\documentclass[10pt]{article}
\usepackage{amsmath}
\usepackage{amssymb}

\usepackage{graphicx}

\usepackage{float}

\usepackage{stmaryrd}

\usepackage{enumitem}

\usepackage{cite}

\usepackage{color} 

\usepackage[labelfont=bf,labelsep=period,justification=raggedright]{caption}

\makeatletter
\renewcommand{\@biblabel}[1]{\quad#1.}
\makeatother

\date{}

\begin{document}

\begin{flushleft}
{\Large
\textbf{The consequences of hesitation: Axelrod model with intrinsic noise}
}
\\
Rémi Perrier$^{\ast}$, 
Yérali Gandica$^{\dagger}$, 
Laura Hern\'{a}ndez$^{\ddagger}$
\\
Laboratoire de Physique Th\'{e}orique et Mod\'{e}lisation, UMR-8089 CNRS, CY Cergy Paris Universit\'{e}, France.
\\
Emails: $\ast$ remi.perrier@cyu.fr, $\dagger$ ygandica@gmail.com, $\ddagger$ laura.hernandez@cyu.fr.
\end{flushleft}
\section*{Abstract}
We study the influence of hesitating agents in the Axelrod model by introducing an \textit{intrinsic noise}, which is proportional to the disagreement between the interacting agents, and thus coupled to the dynamics. Our results show that, unlike the effect of the \textit{cultural drift} where the noise is just controlled by an external parameter, the intrinsic noise never allows the system to reach a frozen state in the thermodynamic limit. Moreover, we show that when the intrinsic noise affects the agents' variables, the system's behaviour is also different from the case when it affects the network of their interactions.

\section{Introduction}\label{intro}

The key contribution of Axelrod's seminal paper on cultural dynamics~\cite{Axelrod1997} was to explain how cultural diversity in a given society, may persist in a model that integrates two well-known aspects of social interactions, namely \textit{homophily} and \textit{social influence}. These terms refer to the tendency of individuals to interact preferentially with like-minded ones and to the fact that their similarity increases with repeated interactions, respectively. As these two characteristics of social interactions tend to homogenize the interacting agents, the cultural diversity observed in society was a paradoxical fact that Axelrod explained as a systemic effect that takes place for some values of the model's parameters. 

The original model consists of a set of agents placed on the sites of a square lattice, where only first neighbours are allowed to interact. Each agent is characterized by a dynamical, F-dimensional, \textit{cultural vector}, representing the agent's cultural attributes, called \textit{features}, like religion, political orientation, preferred sports, etc. For simplicity, each component of this vector takes its value on the same set of $Q$ possible values, called \textit{traits}. 

The probability of interaction is indexed on the overlap between features of the cultural vectors of the interacting agents. In this way, two agents with high overlap are more likely to interact than two agents with a smaller one. During the interaction, the active agent copies the value of one feature chosen at random from the neighbour's cultural vector. Axelrod's dynamics always leads to frozen configurations (meaning that the system does not evolve anymore) which, according to the value of the parameters, go from a culturally homogeneous society to a multicultural one. 

Following the suggestions given by Axelrod himself \cite{Axelrod1997}, several works have introduced variants to the model with the aim to better capture some aspects of social interactions in real life, while still keeping the simplicity and the elegance of the original model. In this way, the model was studied in more realistic topologies than the square lattice, like the small world and the scale-free networks~\cite{klemm2003nonequilibrium}; the influence of mass media was also studied, by means of an external agent able to interact with all the system~\cite{gonzalez2005nonequilibrium,rodriguez2010effects,gandica2011cluster}. In the original model, the trait values are just labels that only indicate whether two agents carry the same value of a given feature, a natural extension is the \textit{metric Axelrod model}, introduced in order to weight the probability of interaction of any two agents by the difference in their traits~\cite{de2009effects,flache2006sustains}. More recently, the symmetry among the features was broken by imposing that one feature could take only two possible values, while the others could take any of the $Q$ possible traits, leading to an induced polarization which was enough to destroy the phase transition in the thermodynamic limit~\cite{gracia2021polarization}.

A largely studied variant of the Axelrod model involves the integration of noise to the dynamics. The spontaneous variation of an agent's state, independently of its state and of those of its neighbours, is known as \textit{cultural drift}. Klemm et al.~\cite{klemm2003global} introduced an external parameter that controls the probability that agents change spontaneously the value of one of their traits. This model leads to a homogeneous state for small noise rates, regardless of the value of $Q$, washing out the phase transition from order to disorder at some critical value of the variability that depends on the number of features, noted $Q_c(F)$. This result is due to the key role of the noise in unblocking frozen configurations, then allowing the Axelrod dynamics to run until the system becomes homogeneous. On the contrary, for large values of noise, it is the ordered state which becomes unstable. For such high values, the noise destroys any homogeneous region. However, the latter is shown to be a size effect, as they observed that the critical value of noise that toggles between these two regimes goes to zero in the thermodynamic limit. 

The fragility of the Axelrod model in the presence of cultural drift led to the search for modifications that could admit a certain amount of external noise and still be robust. For example, Flache and Macy ~\cite{flache2007local} went beyond binary interactions and considered the influence of the whole neighbourhood on the active agent. De Sanctis and Galla studied the effect of setting a threshold for the interaction, coupled with the possibility to overcome it by an \textit{interaction noise}, also controlled by an external parameter~\cite{Sanctis2009} in both, metric and nominal, versions of the Axelrod model. They found that while the external noise reproduces the previous results qualitatively, the interaction noise favours order and does not qualitatively modify the structure of the original model. 

In this article, we study a model where the noise is intrinsically coupled to the dynamics. Unlike cultural drift, where the spontaneous reversal of an agent's feature depends only on an external parameter, here, neighbouring agents may be inclined to revert an already agreed feature if they differ in many others.  The fact that here, the intensity of this disagreement depends on the state variables of the interacting agents, is motivated by the \textit{confirmation bias} principle \cite{Nickerson1998}, which states that in order to form their opinions, people are inclined to choose the heuristics that comforts their own beliefs. In this way, an agent that finds that it agrees with a neighbour in one feature but disagrees in many others, will feel disturbed by this and may choose to destroy that previous agreement in a further step. The larger their dissimilarity (the smaller the overlap between their cultural vectors), the most likely the active agent is to destroy an existing agreement in a particular feature.

This modification of the dynamics removes a  
very strong hypothesis of the original model: if two agents have agreed on a feature at some point of the dynamics, this agreement will be fixed forever, which as said above,  is even less pertinent when the two agents agree on only a few features. 

Some studies integrated the idea of a dynamical noise in order to couple the dynamics of nodes with that of the network. For example, in~\cite{Vazquez2007} {Vazquez et al. consider that} agents are allowed to disconnect from incompatible neighbours (those with whom they differ in all the features) and reconnect to a randomly chosen agent. Their results show that homogeneity is reached for larger $Q$ values than for the standard Axelrod model. Moreover, for high $Q$ values, a \textit{dynamical phase} where the system never freezes is observed. Gracia Lazaro et al.~\cite{GraciaLazaro2011} generalized the model by considering that two agents may disconnect if the fraction of features where they differ exceeds the value of an external parameter representing the tolerance level of the society. The phase diagram shows a non-monotonic dependence of the critical value of the traits $Q_c$ with the tolerance threshold. Specifically, they show that both, very tolerant but also very intolerant societies, reach consensus at high $Q_c$. However in the latter, homogenization is reached only after a long transient where the system remains in a fragmented state. 

Nonetheless, considering that agents are likely to destroy a social link just due to an occasional disagreement is as strong as considering that an agreement on a particular subject will last forever once it had occurred. Therefore, here we study a modification of the Axelrod's model that situates in between these two extreme simplifications by modelling agents that are allowed to \textit{hesitate}.
Unlike cultural drift models, we couple the hesitation probability to the dynamics, under the rationale that an agent is more likely to revise a previous agreement with a neighbour, about a particular feature, if they still disagree in many others. 

The idea of turning back into disagreement a feature where two agents had agreed at a previous stage has been considered {by Radillo et al.} in a deterministic way~\cite{radillo2009axelrod}. In that study, at each time step, agents either undergo the normal Axelrod dynamics or randomly ``undo'' a previous agreement, controlled by an external parameter that sets a threshold on the overlap. The process is deterministic, in the sense that given that the overlap is higher or lower than the threshold, they will either agree or disagree on a previously agreed feature, respectively. In a different approach, here we consider that the disagreement probability is linked to the agents' states. In the case that a feature had already been agreed upon, there is a possibility of reverting this agreement with a probability that is coupled to the agents' overlap. As this probability of destroying a previous agreement is dynamical and coupled to the agents' states, we call it \textit{intrinsic noise}. The relative weight of the agreement (standard Axelrod) and disagreement procedures is modulated by a sensitivity parameter. 

Our results suggest a non-monotonous behaviour of the freezing time as a function of the noise intensity and the traits' variability. Furthermore, we show that the ordered phase of the Axelrod model is unstable to infinitesimal amounts of the intrinsic noise in the thermodynamic limit, as the system enters in a dynamical phase like the one observed in~\cite{Vazquez2007}. 

\section{Model \& Methods}\label{model}
We study a population of N agents located in a square lattice of linear size $L$ with periodic boundary conditions. Each agent $i$, is characterized by a cultural vector, $\overrightarrow \sigma_i$, of dimension $F$, where each component codes for a \textit{cultural feature}, that can take any integer value, known as \textit{cultural trait}, between one and $Q$: $\sigma_i^f \in$ $\{1, ..., Q\}$, with $f \in \{1, ..., F\}$.
These features aim to model the agent’s cultural attributes (language, religion, political orientation, etc.). Such a model is agnostic regarding the traits as their value do not carry any specific meaning.\\

The time-discrete dynamics is defined by the following steps:
\begin{itemize}
	\item An agent $i$ (the \textit{active agent}) and one of its four nearest neighbors, $j$, are randomly chosen.
	\item The \textit{overlap} between their vectors is computed as:
	
	$\ell(i,j) = \sum_{f=1}^{F}\delta(\sigma_i^f, \sigma_j^f)$, with $\delta(\sigma_i^f, \sigma_j^f)$ being the Kronecker delta
	
If $0<\ell(i,j)<F$ ( the two sites neither fully disagree nor fully agree), the bond $(i,j)$ is said to be \textit{active}, otherwise \textit{inactive}.
	 \begin{itemize}
	 \item If the bond is inactive, the procedure is restarted by choosing another active site.
	 \item If the bond is active, a component of the cultural vector of the active agent, $\sigma_i^f$, is randomly chosen with probability {$\frac{1}{F}$}. It should be noticed that this step differs from the original Axelrod model (where the component would be chosen at random within the set of features having \textit{different values} for the two agents $\mathcal{G} = \{k\in [1,F]|\sigma_i^k \neq \sigma_j^k\}$). We denote this modification, the \textit{full set variant}. 
	 
	 Two different interaction processes can take place at this point: agreement (as per the original Axelrod model) or disagreement (intrinsic noise driven).
	 \begin{enumerate}
	 \item Agreement: if $\sigma_i^f \neq \sigma_j^f$ (the two agents disagree on the value for this feature), then, the usual Axelrod procedure takes place. The agents interact with probability {$\omega_{i,j} = \frac{\ell(i,j)}{F}$} and the interaction consists in setting the feature of the active agent
	 to the neighbor's trait value. 
	 
	 \item Disagreement: due to the full set variant used here, a choice where $\sigma_i^f = \sigma_j^f$ (the two agents already agree on the value for this feature) is possible, then, the active agent may revert the agreement previously obtained with a probability proportional to the disagreement with the neighbour, $\chi\times\left(1-\omega_{i,j}\right)$, where the proportionality constant $\chi \in \left[0,1\right]$ modulates the intensity of the disagreement. By disagreement, we mean that the active agent replaces the component $\sigma_i^f$ by a trait randomly chosen between $1$ and $Q$, with uniform probability. Notice that, once the coincident feature has been chosen,  the greater the number of different components, the higher the probability that the active agent destroys a previous agreement. This probability is called \textit{intrinsic noise} because it is coupled to the dynamics of the system, instead of being determined only by an external parameter as in the \textit{cultural drift} model\cite{klemm2003global}. 
	 \end{enumerate}
	 
	 \end{itemize}
\end{itemize}

The idea behind the disagreement procedure is to model a \textit{hesitation} process that occurs when an agent observes that it agrees in a feature with a neighbour with whom it disagrees in many others (low overlap $\ell(i,j)$) between them. Then, when interacting with this neighbour again, the active agent may question the  previous agreement  and eventually revert it.
This can be seen as a corollary to the confirmation bias, where the agent chooses not to keep an opinion that links it to  an  "information environment" that is drastically different from its beliefs (interacting with a neighbour with a low overlap), thus reinforcing the difference.
 

The initial states of each trait are drawn from a uniform distribution : $\sigma_i^f(0) \in \mathcal{U}\llbracket1,Q\rrbracket \text{ }\forall i,f$. The algorithm described above is iterated until there is no active link left, in which case the system is said to be in a \textit{frozen state}: $\ell(i,j) = \{0,F\}\text{ }\forall i,j$. 
In general, several system sizes are studied and the results presented here correspond to averages over $100$ independent realizations, with the exception of the phase plots where carrying out such extensive computations for all the considered points of the phase space was unfeasible and where the results have been averaged over  $10$ realizations. These plots   are only used to roughly locate the interesting regions to be explored in larger detail. 

We compute the mean normalized size of the largest cluster $\langle \frac{S_\mathrm{max}}{N} \rangle$, the average freezing time (or average convergence time) $\langle T \rangle$ and we will subsequently use the average freezing time per site as well, that we denote $\langle \tilde{\tau} \rangle$ (total number of iterations required to reach the frozen state divided by the system size).

We observe that, unlike in models with parametric external noise, for certain values of the parameters $(N,F,Q,\chi)$, the system might not reach the frozen state in a reasonable number of iterations, if at all. This observation leads us to set an upper threshold for the number of iterations. If this upper threshold is reached, the algorithm is stopped. Here it is set at $10^9$ total iterations, {which is well above both the convergence times usually found in previous works and the convergence times we observe in our model with zero noise ($\sim 10^7$ in the frozen order region and $\sim 10^6$ in the frozen disorder region)}, while still achievable in reasonable CPU time. As a consequence, this leads to a size-dependent threshold in terms of average freezing time per site.

\section{Results}

\begin{figure}[H]
	\centering
	\includegraphics[width=\linewidth]{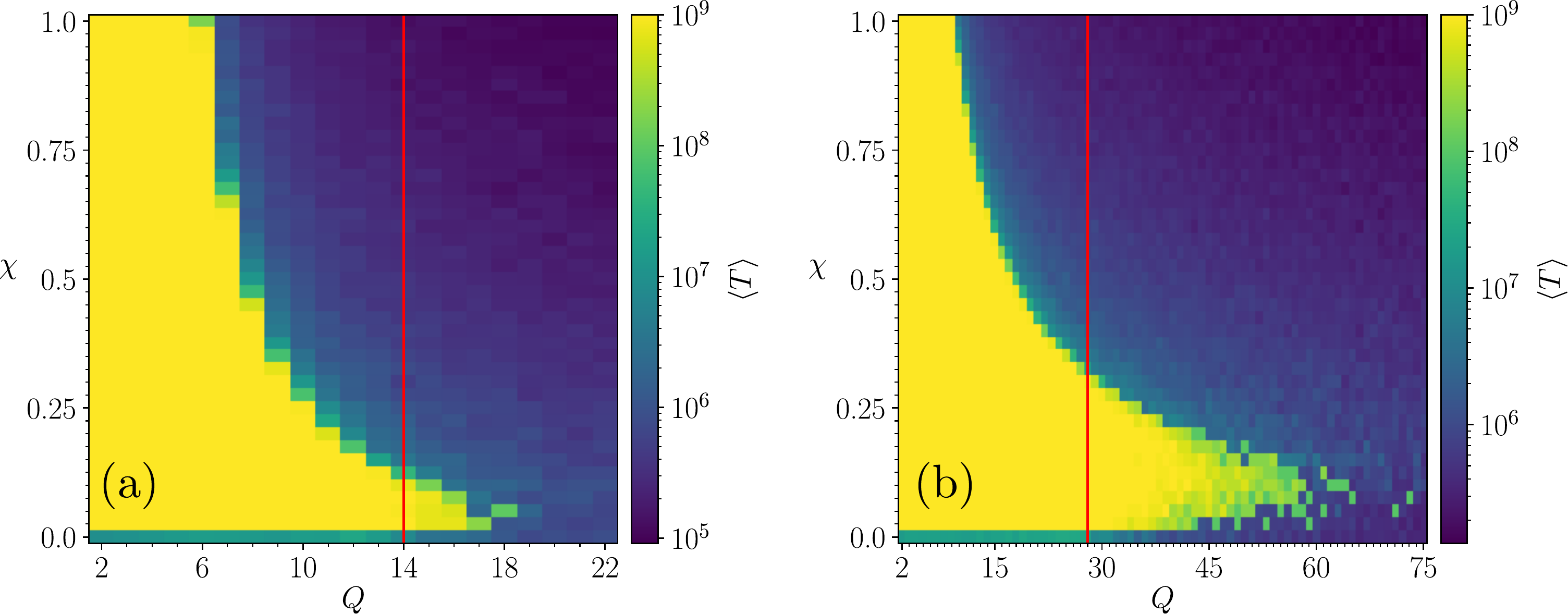}
	\caption{Number of total iterations for different values of $Q$, and $\chi$, for systems of size $N=30^2$ agents. 
	The algorithm is stopped when it reaches $10^9$ iterations. Red lines correspond to $Q_c^0$, the value of the critical point in the original Axelrod model ($Q_c^0 \approx 14$ for $F=3$ and $Q_c^0 \approx 28$ for $F=5$). Each point corresponds to an average over $10$ realizations: (a) $F=3$ ($861$ points) and (b) $F=5$ ($3034$ points).}
	\label{F3_F5_iter}
\end{figure}

We have observed that when the intrinsic noise has the same weight as the agreement probability in the standard Axelrod procedure (with probability $(1-\omega_{i,j})$ and $\omega_{i,j}$ respectively), then the system never converges for low $Q$ values. Therefore, we have introduced a modulation to the intrinsic noise in the form of an external parameter $\chi \in [0,1]$ that controls its intensity, and we started our study by exploring how it affects the convergence time. In Figure \ref{F3_F5_iter}, we show the number of iterations as a function of $Q$ and $\chi$, for $F=3$ in (a) and $F=5$ in (b). Here, we need to remind that when the iterations reach $10^9$ it means that the dynamics was stopped before convergence. {Because of the long computation time involved, these exploratory phase plots were computed over an sample of only 10 realizations, and have fluctuations of the order of 20\%. Nevertheless, these plots are still useful to locate the region of interest to be studied in larger detail which is described below.}

Fig \ref{F3_F5_iter} reveals three important features: \begin{enumerate}[nosep,label=\alph*)]
\item The existence of intrinsic noise prevents freezing in the region of $Q$ where the standard Axelrod model reaches frozen order.
\item A non-monotonous behaviour of the convergence time is observed as a function of $\chi$ and $Q$, with a re-entrant region of divergent times for high $Q$ and relatively low noise. 
\item The re-entrant region of divergent times, mentioned in the previous point, is located in the region where $Q$ is high enough so that the original model would have reached a frozen disordered state. 
\end{enumerate}

\begin{figure}[H]
	\centering
	\includegraphics[width=\linewidth]{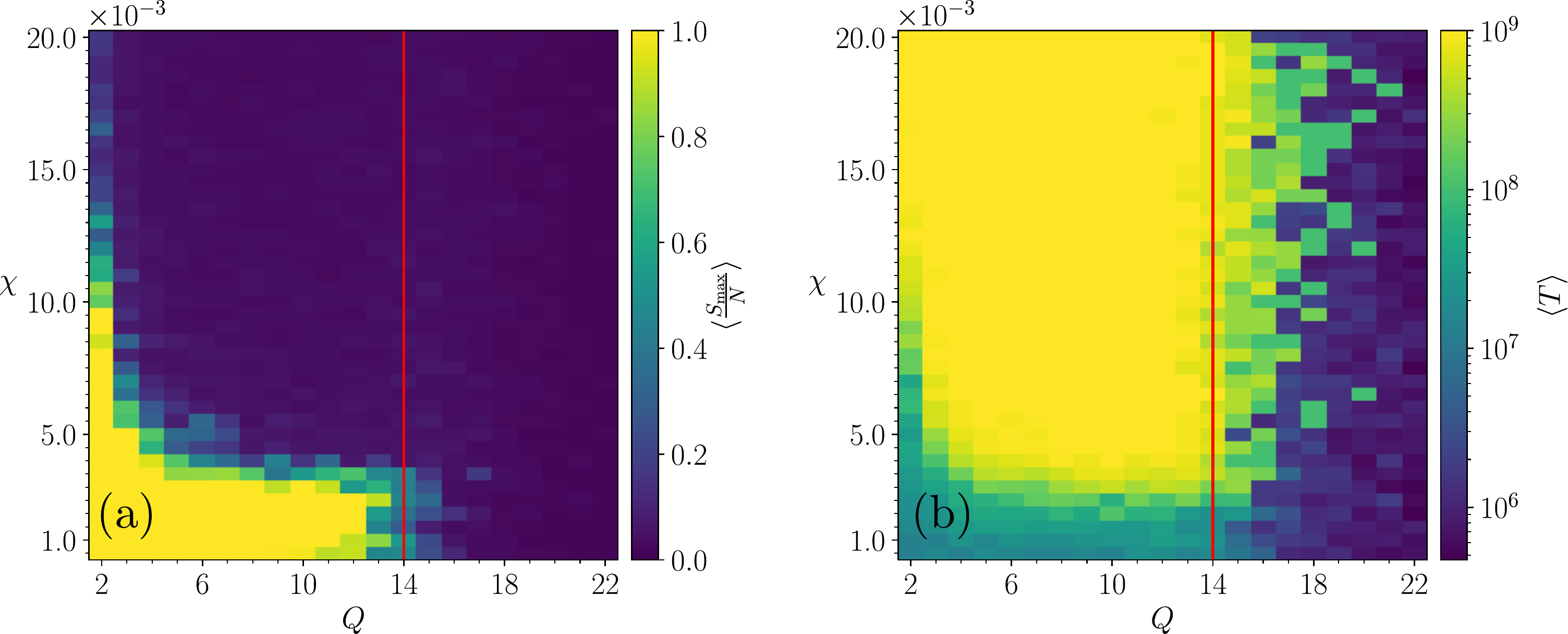}
	\caption{Zoom on the phase plot in the $(Q, \chi)$ plane for very low sensitivity values. Each of the $840$ points corresponds to an average over $10$ realizations, for a system of $N=30^2$, $F=3$ at fixed $Q$ and $\chi$. Red lines correspond to $Q_c^0$, the value of the critical point in the original Axelrod model ($Q_c^0 \approx 14$ for $F=3$). (a) Mean relative size of the largest cluster. (b) Freezing time.}
	\label{F3_iter_Smax}
\end{figure}
\newpage
Let us consider now the region for $Q<Q_c^0$, where $Q_c^0$ defines the order-disorder transition point on the standard Axelrod model with zero noise. This is shown in Fig. \ref{F3_F5_iter} by a green horizontal single-pixel-wide line at the bottom of the plots, where the system freezes in {$\sim 10^7$ iterations}. Figure \ref{F3_iter_Smax} shows a zoom of Fig.~\ref{F3_F5_iter} which reveals the details of the behaviour of both the convergence time and the order parameter, $\langle \frac{S_\mathrm{max}}{N} \rangle$, in the region of extremely low intensity of the intrinsic noise.

\begin{figure}[H]
	\centering
	\includegraphics[width=.3\linewidth]{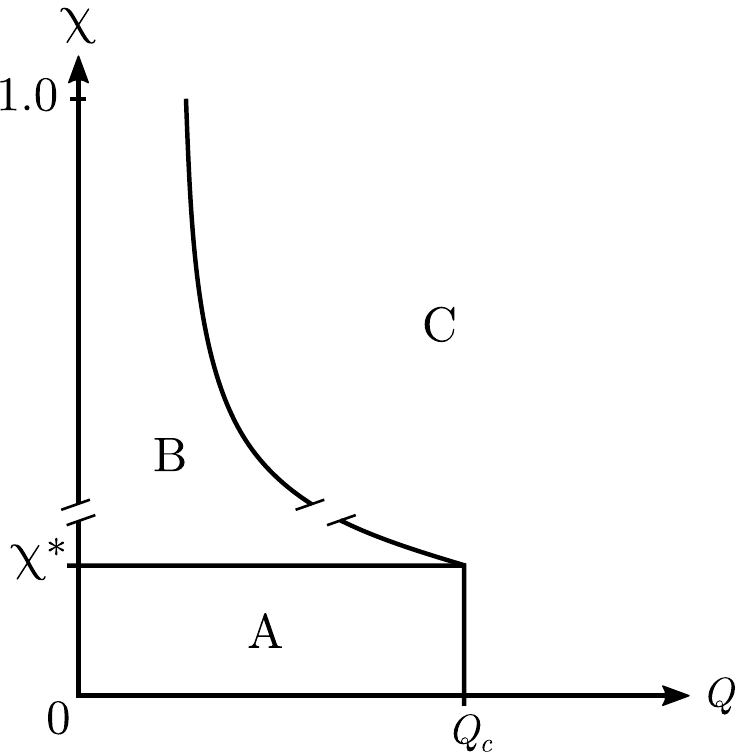}
	\caption{Schematic phase diagram: Regions of the $(Q, \chi)$ plane showing qualitatively different behaviors. Lines are a only a guide to the eye; note the line interruptions that show that the graph is out of scale (for example for $N = 40^2$ and $F=3$, $\chi^* = 2.2\times10^{-3}$ and $Q_c = 14$). A: Frozen order; B: Non-frozen dynamical phase; C: Frozen disorder. }
	\label{phase_diagram}
\end{figure}

Figure~\ref{phase_diagram} schematically shows the regions of qualitatively different behaviour, observed Fig.\ref{F3_F5_iter}, and Fig.\ref{F3_iter_Smax}. We can see three regions: A: Frozen order; B: Non-frozen dynamical phase; C: Frozen disorder.

\begin{figure}[H]
	\centering
	\includegraphics[width=.53\linewidth]{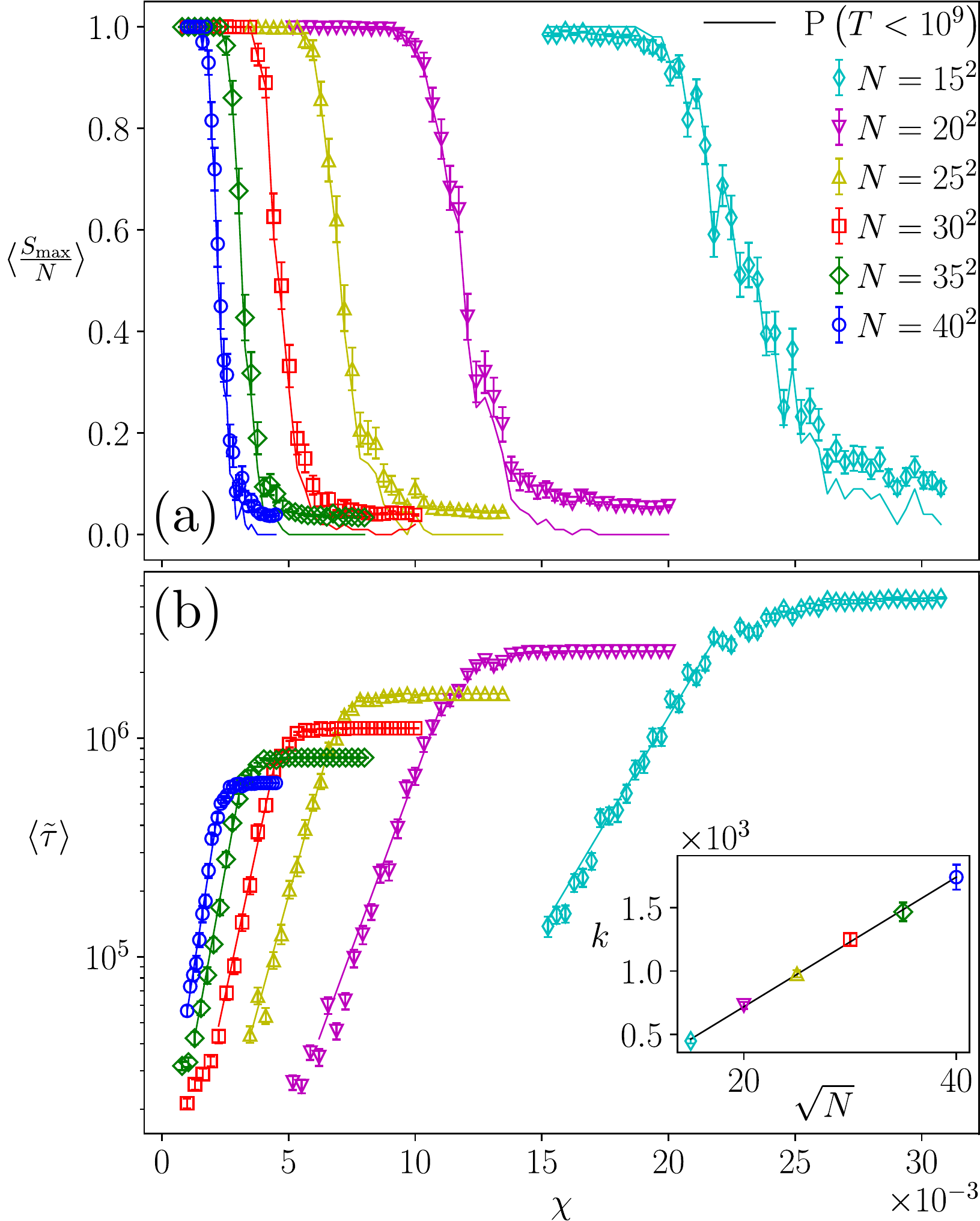}
	\caption{Crossing the non-freezing point, $F=3$ $Q=8$, average over $100$ realizations. (a) Mean relative size of the largest cluster (symbols), and  $P\left(T<10^9\right)$, the probability of reaching the frozen state for each size (solid lines). (b) Average freezing time per site (threshold: {$\frac{10^9}{N}$}). Lines are the fits $\langle \tilde{\tau} \rangle = e^{k(N)\chi+b}$. Inset: Fitted coefficient $k(N)$ versus $\sqrt{N}$, line is the fit $k(N) = c\sqrt{N}+d$.}
	\label{F3_Smax_tau}
\end{figure}

 Figure~\ref{F3_Smax_tau} shows the behaviour of the order parameter and the freezing time in the $Q<Q_c^0$ region as a function of the noise intensity for various system sizes. 
 This corresponds to crossing from region A to region B in Fig. \ref{phase_diagram} at a fixed $Q<Q_c^0$, for a given $F$. We have found that at any N, the freezing time per site grows with an exponential dependence of the sensitivity. Specifically, $\langle \tilde{\tau} \rangle = e^{k(N)\chi+b}$, where the slope of the exponent behaves as $k(N) \propto \sqrt{N} $. This leads to a divergent asymptotic behaviour of the freezing time per site in the thermodynamic limit. In other words: any finite amount of intrinsic noise results in an infinite freezing time per site as $N \rightarrow \infty$, preventing the system from reaching its (ordered) steady state. 
 The horizontal part of all the $\langle \tilde{\tau} \rangle$ as a function of $\chi$ curves, observed above a given value of sensitivity $\chi^*$ (which is a function of $N$ and the cutoff time) in Fig.~\ref{F3_Smax_tau}, is just an artefact of the upper cutoff time imposed on our simulations, as detailed in Section \ref{model}, and indicates that the system does not freeze in that region.
 
 As the dynamics is stopped while the system is still in a non-frozen state, the order parameter is systematically low (i.e. no clusters could grow although the maximum allowed time was reached). This causes $\langle\frac{S_\mathrm{max}}{N}\rangle$ averaged over different realizations to decrease, indicating that, as $\chi$ increases, more and more samples are unable to reach a frozen ordered state. This can be seen in Fig.~\ref{F3_Smax_tau}(a) where we have plotted (full line) the probability that the number of iterations $T$ is smaller than the limit of $10^9$ which superposes with the  $\langle\frac{S_\mathrm{max}}{N}\rangle$  curves. In this sense, Fig.~\ref{F3_Smax_tau}(a) should not be interpreted as an order-disorder transition but as a transition between the frozen order and the "dynamical phase" where the system continually re-configures without reaching any stationary state.

\begin{figure}[H]
	\centering
	\includegraphics[width=.53\linewidth]{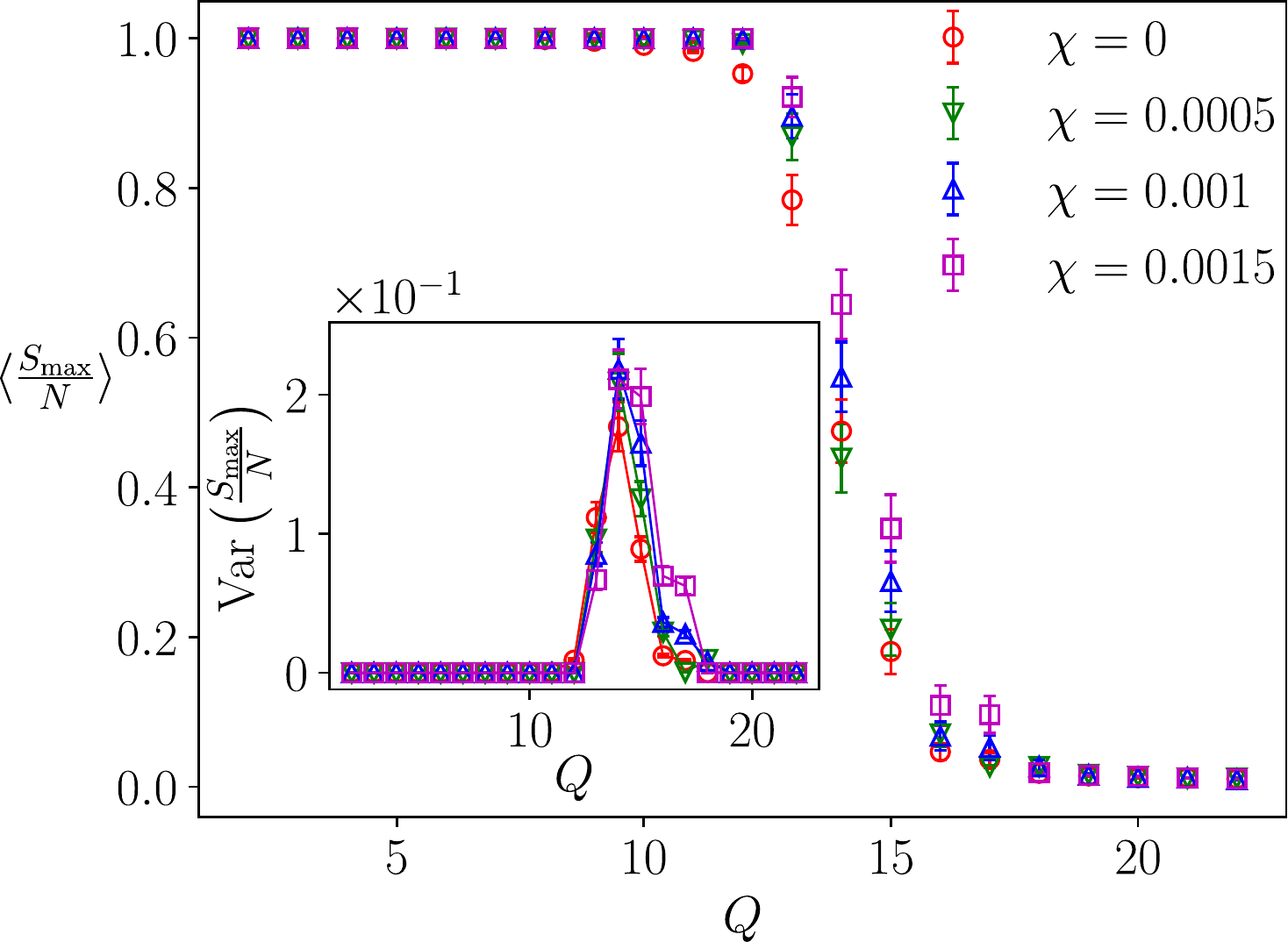}
	\caption{Order-disorder transition with non-disruptive level of noise (low $\chi$ values such that the finite system converges), $N=40^2$, $F=3$, average over $100$ realizations (threshold : $10^9$ iterations). }
	\label{Smax_F3_smallchi}
\end{figure}

In order to investigate whether the addition of low-intensity intrinsic noise modifies the transition of the standard Axelrod model, we study a horizontal crossing from region A to region C in Fig \ref{phase_diagram}, at various fixed $\chi<\chi^*$, for a given $F$ and $N$. Fig \ref{Smax_F3_smallchi} shows the order parameter as a function of $Q$ for different noise intensities, below the value $\chi^*$ that prevents convergence. For the shown case of $F=3$ and $N=40^2$, the system reaches the maximum allowed number of iterations for sensitivities above $\chi^* = 2.2\times10^{-3}$. If
\noindent
we keep the sensitivity below this critical level of noise, we can observe the usual frozen order to frozen disorder transition. It appears that the associated critical value $Q_c$ is not affected by these non-disruptive levels of noise, as can be seen in the inset of Fig.~\ref{Smax_F3_smallchi} where the maximum of the variance is located at the same value, $Q_c=Q_c^0=14$.

\section{Discussion}

We have investigated the role of hesitation in the Axelrod model by introducing noise coupled to the dynamics. The effect of this intrinsic noise accounts for the fact that agents may destroy a partial agreement previously obtained with some of their neighbours with a dynamical probability that depends on the overlap of their vectors. In this way, an active agent who has to reconsider a feature where it had previously agreed with its neighbour, will be more prone to revise it, if they still disagree on many others.

In general, we found that {with the exception of the region of the phase diagram with either very large Q with respect to the $Q_c$ of the standard Axelrod model, or  high noise intensity, where the system trivially freezes in a  disordered phase}, 
 the absorbent states of the original Axelrod model are not robust face to the introduction of intrinsic noise, 
 as even relatively small amounts of it prevent the system from converging to a steady state, leading to a \textit{dynamical phase} where the system keeps trying to grow clusters of similar agents, that get continuously destroyed, even for $Q<Q_c$. This result is very different from the results of the Axelrod model with \textit{external parametric} noise~\cite{klemm2003global,flache2007local,Sanctis2009}. 

We also show that whether the intrinsic noise affects the node variables or the interaction network topology is determinant for the outcomes of the dynamics. Unlike previous works~\cite{Vazquez2007,GraciaLazaro2011}, where the dynamical noise affects the topology of interactions, letting the cultural vectors unchanged, here, on the contrary, the dynamical noise affects the agents' cultural vectors, letting the topology unchanged, and the two dynamics lead the system through different paths. When the intrinsic noise allows for a modification of the network topology through the rewiring of the connections, a dynamical (perpetual) phase is observed only for large $Q$ values~\cite{Vazquez2007}. For smaller $Q$ values, but still leading to fragmentation in the original model, the system can reach order, i.e., homogeneous states, by a mechanism similar to the one behind the external noise applied to agents’ variables ~\cite{klemm2003global, flache2007local, Sanctis2009}. In this work, we are facing a different scenario. We show that when the intrinsic noise affects the agents' values, then the system enters a dynamical phase even for low $Q$, as soon as the noise intensity is above a very low threshold, which is a finite-size effect that disappears in the thermodynamic limit. In other words, allowing the agents to revert a previous agreement when they have a low overlap leads to a dynamical phase in the region of low and moderate  $Q$ values, while allowing the agents to rewire their connections does not, except for very high $Q$ values. 

For extremely low values of the noise intensity, where the system converges, we observe that the transition occurs at the same $Q_c$ as the original model (without noise). However, strong finite-size effects are observed, and this region (region A of Fig.~\ref{phase_diagram}) seems to vanish in the thermodynamic limit, where an infinitesimal amount of intrinsic noise leads the system to the dynamical phase. 
 
Finite-size effects affecting the convergence times go exponentially with the noise intensity as $\langle \tilde{\tau} \rangle = e^{k(N)\chi+b}$ with the coefficient $k \propto \sqrt{N}$. The fact that the active agent can disagree with a neighbour, in a previously agreed feature, prevents the deactivation of links, and therefore the system does not reach the frozen state. 
 Surprisingly, even the extremely low probabilities given by low values of the intensity $\chi$ are still enough to prevent both the homogeneous and the fragmented phases to set in.

In conclusion, it is interesting to notice that in the thermodynamic limit, a finite but fixed amount of external noise helps the Axelrod model to find its ordered state, removing the barriers imposed by the strong assumption of non-interaction in case of zero overlapping. However, any finite intrinsic noise proportional to the non-overlapping keeps the system exploring its states space. The initial homogeneous disorder never allows the system to grow large enough clusters. This is interesting, above all, if we think that society is always in a dynamic state, and although it is fragmented into clusters, actually disagreements are conditioned by our interactions rather than by random fluctuations.

\section*{Acknowledgments}
The authors acknowledge the OpLaDyn grant obtained in the 4th round of the Trans-Atlantic Platform Digging into Data Challenge (2016-147 ANR OPLADYN TAP-DD2016) and Labex MME-DII (Grant No. ANR reference 11-LABEX-0023).


\begin{thebibliography}{10}
\providecommand{\url}[1]{\texttt{#1}}
\providecommand{\urlprefix}{URL }
\expandafter\ifx\csname urlstyle\endcsname\relax
  \providecommand{\doi}[1]{doi:\discretionary{}{}{}#1}\else
  \providecommand{\doi}{doi:\discretionary{}{}{}\begingroup
  \urlstyle{rm}\Url}\fi
\providecommand{\bibAnnoteFile}[1]{%
  \IfFileExists{#1}{\begin{quotation}\noindent\textsc{Key:} #1\\
  \textsc{Annotation:}\ \input{#1}\end{quotation}}{}}
\providecommand{\bibAnnote}[2]{%
  \begin{quotation}\noindent\textsc{Key:} #1\\
  \textsc{Annotation:}\ #2\end{quotation}}
\providecommand{\eprint}[2][]{\url{#2}}

\bibitem{Axelrod1997}
Axelrod R (1997) The dissemination of culture.
\newblock Journal of Conflict Resolution 41: 203--226.
\input{Axelrod1997}

\bibitem{klemm2003nonequilibrium}
Klemm K, Egu{\'i}luz VM, Toral R, San~Miguel M (2003) Nonequilibrium
  transitions in complex networks: A model of social interaction.
\newblock Physical review E 67: 026120.
\input{klemm2003nonequilibrium}

\bibitem{gonzalez2005nonequilibrium}
Gonz{\'a}lez-Avella JC, Cosenza MG, Tucci K (2005) Nonequilibrium transition
  induced by mass media in a model for social influence.
\newblock Physical Review E 72: 065102.
\input{gonzalez2005nonequilibrium}

\bibitem{rodriguez2010effects}
Rodr{\'i}guez AH, Moreno Y (2010) Effects of mass media action on the axelrod
  model with social influence.
\newblock Physical Review E 82: 016111.
\input{rodriguez2010effects}

\bibitem{gandica2011cluster}
Gandica Y, Charmell A, Villegas-Febres J, Bonalde I (2011) Cluster-size entropy
  in the axelrod model of social influence: Small-world networks and mass
  media.
\newblock Physical Review E 84: 046109.
\input{gandica2011cluster}

\bibitem{de2009effects}
De~Sanctis L, Galla T (2009) Effects of noise and confidence thresholds in
  nominal and metric axelrod dynamics of social influence.
\newblock Physical Review E 79: 046108.
\input{de2009effects}

\bibitem{flache2006sustains}
Flache A, Macy MW (2006) What sustains cultural diversity and what undermines
  it? axelrod and beyond.
\newblock arXiv preprint physics/0604201 .
\input{flache2006sustains}

\bibitem{gracia2021polarization}
Gracia-L{\'a}zaro C, Brigatti E, Hern{\'a}ndez AR, Moreno Y (2021) Polarization
  inhibits the phase transition of axelrod's model.
\newblock arXiv preprint arXiv:210206921 .
\input{gracia2021polarization}

\bibitem{klemm2003global}
Klemm K, Egu{\'\i}luz VM, Toral R, San~Miguel M (2003) Global culture: A
  noise-induced transition in finite systems.
\newblock Physical Review E 67: 045101.
\input{klemm2003global}

\bibitem{flache2007local}
Flache A, Macy MW (2007) Local convergence and global diversity: the robustness
  of cultural homophily.
\newblock arXiv preprint physics/0701333 .
\input{flache2007local}

\bibitem{Sanctis2009}
Sanctis LD, Galla T (2009) Effects of noise and confidence thresholds in
  nominal and metric axelrod dynamics of social influence.
\newblock Physical Review E 79.
\input{Sanctis2009}

\bibitem{Nickerson1998}
Nickerson RS (1998) Confirmation bias: A ubiquitous phenomenon in many guises.
\newblock Review of General Psychology 2: 175--220.
\input{Nickerson1998}

\bibitem{Vazquez2007}
Vazquez F, Gonz{\'a}lez-Avella JC, Egu{\'i}luz VM, Miguel MS (2007) Time-scale
  competition leading to fragmentation and recombination transitions in the
  coevolution of network and states.
\newblock Physical Review E 76.
\input{Vazquez2007}

\bibitem{GraciaLazaro2011}
Gracia-L{\'a}zaro C, Quijandr{\'i}a F, Hern{\'a}ndez L, Flor{\'i}a LM, Moreno Y
  (2011) Coevolutionary network approach to cultural dynamics controlled by
  intolerance.
\newblock Physical Review E 84.
\input{GraciaLazaro2011}

\bibitem{radillo2009axelrod}
Radillo-D{\'i}az A, P{\'e}rez LA, del Castillo-Mussot M (2009) Axelrod models
  of social influence with cultural repulsion.
\newblock Physical Review E 80: 066107.
\input{radillo2009axelrod}

\end{thebibliography}
\end{document}